\documentclass{article}

\usepackage{graphicx} 
\usepackage[utf8]{inputenc}
\usepackage{palatino}
\usepackage{amsmath}
\usepackage{amssymb}
\usepackage{graphicx}
\usepackage[labelfont=bf]{caption}
\usepackage{subcaption}
\usepackage{setspace}

\usepackage[margin=1in, headsep=1cm, bottom=5cm]{geometry}

\usepackage[hidelinks]{hyperref}
\usepackage{tabu}
\usepackage{cite}
\usepackage[table]{xcolor}
\usepackage{nomencl}
\usepackage[nonumberlist,nogroupskip,xindy]{glossaries}
\usepackage{floatrow}
\usepackage{wrapfig}

\usepackage{fancyhdr}
\usepackage{lmodern}
\usepackage{comment}
\usepackage{multirow}
\usepackage{rotating}
\usepackage[sectionbib]{bibunits}

\usepackage{etoolbox}

\defaultbibliographystyle{unsrt} 
\defaultbibliography{references}

\setglossarystyle{long}

\newcommand*{\email}[1]{%
    \normalsize\href{mailto:#1}{#1}\par
    }

\newtoggle{mrm}

\togglefalse{mrm}

\newcommand{\mytitle}{Phase-sensitive modeling improves Fat DESPOT multiparametric relaxation mapping in fat-water mixtures}

\begin{document}

\begin{spacing}{1}
\begin{center}

\textbf{\Large{\mytitle}}
\newline
\newline

\large{Renée-Claude Bider$^1$, Cristian Ciobanu$^1$, Jorge Campos Pazmiño$^{1,2}$, Véronique Fortier$^{1,3,4,5}$, Evan McNabb$^{3,4}$, Ives R. Levesque$^{1,2,4,5,6}$*}
\end{center}

\noindent\textbf{1} Medical Physics Unit, McGill University, Montréal, QC, Canada\\ \\
\textbf{2} Department of Physics, McGill University, Montréal, QC, Canada\\ \\
\textbf{3} Medical Imaging, McGill University Health Center, Montréal, QC, Canada\\ \\
\textbf{4} Department of Diagnostic Radiology, McGill University, Montréal, QC, Canada\\ \\
\textbf{5} Gerald Bronfman Department of Oncology, McGill University, Montréal, QC, Canada\\ \\
\textbf{6} Research Institute of the McGill University Health Centre, Montréal, QC, Canada\\ \\
\textbf{*}Corresponding author:\\ \\
\indent\indent
\begin{tabular}{>{\bfseries}rl}
\textbf{Name}&  Ives R. Levesque\\
\textbf{Department}& Medical Physics Unit\\
\textbf{Institute}& McGill University\\
\textbf{Address}&  Cedars Cancer Centre, DS1.9326 1001 boul Décarie, Montréal,\\ 
			& Québec, Canada, H4A 3J1\\
\textbf{E-mail}& \email{ives.levesque@mcgill.ca}\\
\end{tabular}
\\ 
\\
\\
\\
\noindent \today\\
\\ 
\\ 
\\
\noindent \textbf{ Manuscript word count:} 4944\\
\noindent \textbf{Abstract word count:} 249\\
\end{spacing}
\newpage

\section*{Abstract}

\textbf{Purpose}: To improve on the original form of Fat DESPOT, a multiparametric mapping technique that returns the fat- and water-specific estimates of $R_1$ ($R_{1f}$, $R_{1w}$),  $R_{2}^*$, and proton density fat fraction (PDFF) by upgrading the fat-water separation method used for selection of initial parameter guesses, and by introducing explicit model sensitivity to the phase of the water and fat signals.
\\
\textbf{Methods}: We compared the 3-point Dixon and Graph Cut (GC) approaches to initial guesses for Fat DESPOT in phantom experiments at 3 T in a variable fat fraction gel phantom. Also in phantom, we then compared the original Fat DESPOT approach to a magnitude approach modeling the phases of fat and water separately (Fat DESPOT$_{m\phi}$), and an approach that models the complex data (Fat DESPOT$_c$). The best-performing approach was then used in the lower leg of a healthy human participant. 
\\
\textbf{Results}: In phantoms, Fat DESPOT using the 3-point Dixon and GC performed similarly in parametric estimates and precision, though the Dixon approach deviated from the overall trend in the 50\% nominal fat fraction ROI. Furthermore, Fat DESPOT$_c$ showed the best agreement with reference PDFF (average error 1.5$\pm$1.2\%) and the lowest combined standard deviation across ROIs, for PDFF, $R_{1f}$, and $R_{1w}$ ($\sigma$ = 0.13\%, 0.19 s$^{-1}$, 0.0082 s$^{-1}$,).
\\
\textbf{Conclusion}: With a higher precision of $R_{1f}$ and $R_{1w}$, accuracy of PDFF, and more echo time versatility than other compared approaches, this work demonstrates the advantages of the GC approach for initial guesses paired with complex fitting for Fat DESPOT multiparametric imaging.

\noindent \textbf{Keywords:} Multiparametric imaging, Relaxation mapping, Fat-water separation, Fat relaxation rate, Water relaxation rate 

\newpage
\section*{Introduction}

Quantitative MRI mapping of proton density fat fraction (PDFF) and relaxation parameters $R_1$ and $R_2^*$ offers promising insights into disease. Notably, mapping $R_2^*$ and PDFF in the pancreas is sensitive to iron content in patients with a variety of diseases \cite{pfeifer_pancreatic_2015, santarelli_estimation_2018}. Meanwhile, mapping PDFF, $R_1$, and $R_2^*$, could differentiate between types of liver disease\cite{schaapman_multiparametric_2021, banerjee_multiparametric_2014} and correlate with treatment outcomes \cite{pavlides_multiparametric_2016, banerjee_multiparametric_2014, jaubert_multi-parametric_2020}. Finally, $R_2^*$ and $R_1$ mapping is valuable in tumour hypoxia \cite{arai_oxygen-sensitive_2021}. However, when separate acquisition protocols are required for each measured parameter, long acquisition times are taxing on patients, increasing the risk of motion artifacts and limiting dynamic imaging.
Multiparametric mapping, where a single acquisition protocol obtains maps for several parameters, can significantly reduce acquisition times \cite{jaubert_multi-parametric_2020, le_ster_fast_2017}.

The original approach to  fat-water relaxation mapping\cite{le_ster_breath-hold_2016, le_ster_fast_2017}, more recently referred to as fat-water separated driven equilibrium single point observation of $T_{1}$ or "Fat DESPOT"\cite{fortier_mr-oximetry_2023}, is a multiparametric technique that models the signal magnitude in a variable flip angle (VFA) multi echo gradient echo (mGRE) experiment to obtain maps for PDFF, $R_2^*$, $R_1$ of water ($R_{1w}$), and $R_1$ of fat  ($R_{1f}$), simultaneously. Fat DESPOT is versatile in its potential applications. The isolated $R_{1w}$ could be used in MRI-based assessments of liver disease \cite{wan_water_2022, michelotti_empirical_2024}. $R_{1f}$ mapping has been proposed as a sensitive method for $R_1$-based MR oximetry \cite{colliez_qualification_2014} due to the increased solubility of oxygen in fat relative to water. Prior work has demonstrated the oxygenation sensitivity of Fat DESPOT$_m$ in phantoms \cite{fortier_mr-oximetry_2023}. In addition, the 3D mGRE sequence required for Fat DESPOT is broadly available on clinical scanners, enhancing translatability.

The magnitude signal model is referred to in this work as Fat DESPOT$_m$. Building on this, we introduce the phase-sensitive Fat DESPOT$_{m\phi}$ and Fat DESPOT$_c$ models. Like Fat DESPOT$_{m}$, Fat DESPOT$_{m\phi}$ is a magnitude model, but it incorporates a term for the initial phase difference between water and fat signals. Fat DESPOT$_c$ considers the full complex signal. We begin by comparing the use of two fat-water separation techniques, 3-point Dixon and Graph Cut (GC), to provide initial guesses as input to the Fat DESPOT$_{m}$ fit. Then, we conduct a systematic comparison of Fat DESPOT$_m$, Fat DESPOT$_{m\phi}$, and Fat DESPOT$_c$ in phantom, to assess performance across a wide range of fat fractions. Finally, we present a pilot measurement in the lower leg of a human volunteer using the Fat DESPOT$_c$ approach.

\section*{Methods}

All calculations were completed in MATLAB (Mathworks, USA, R2023a).

\subsection*{Signal model}

This paper compares three models for Fat DESPOT mapping of PDFF, $R_2^*$, $R_{1f}$, and $R_{1w}$. The most general approach, Fat DESPOT$_c$ (equation \ref{eq:complexFD_phi0}), models the complex mGRE signal, accounting for the $B_0$ field inhomogeneity ($\Delta B_0$; fixed parameter), and for the initial phases of fat and water magnetization, $\phi_{0f}$ and $\phi_{0w}$, free parameters which are believed to be different \cite{wang_t1_2020}.
Taking the magnitude of the complex model yields a second model, Fat DESPOT$_{m\phi}$ (equation \ref{eq:magphiFD2}), where the initial phase difference between fat and water, $\Delta\phi_0=\phi_{0f}-\phi_{0w}$, remains as a free parameter in the fit.
Finally, Fat DESPOT$_m$ models the magnitude signal and assumes that the initial phases of fat and water magnetization are identical ($\phi_{0f}$=$\phi_{0w}$) such that $\phi_0=0$ drops out of the equation, identical to previous work \cite{fortier_mr-oximetry_2023}.
For all models, $f$ is the PDFF,  $F$ and $W$ are the steady-state fat and water signal components for TE = 0 and normalized by the equilibrium magnetization (equations \ref{eq:F2} and \ref{eq:W2}, respectively), and the fat spectrum with $N$ resonances is represented by amplitudes $A_n$ and frequency shifts $\Delta \omega_n$.
To allow for $B_0$ field map variations between acquisitions with different FA, which we observed experimentally, the parameter $\Delta B_{0,\theta}$ was introduced into the model as a flip-angle specific fixed parameter.

\begin{equation}
    S_{c}(\mathrm{TE}, \mathrm{TR}, \theta)=S_0\left[(1-f)We^{-i\phi_{0w}}+fF\sum_{n=1}^{N} A_ne^{-i\Delta \omega_n \mathrm{TE}}e^{-i\phi_{0f}}\right]e^{-R_2^{*}\mathrm{TE}}e^{-i\gamma\Delta B_{0,\theta}\mathrm{TE}}
    \label{eq:complexFD_phi0}
\end{equation}

\begin{equation}
    S_{m\phi}(\mathrm{TE}, \mathrm{TR}, \theta)=S_{0} e^{-R_2^{*} \mathrm{TE}} \sqrt{\left[(1 - f)W + fF\sum_{n=1}^{N} A_n \cos \left(\Delta \omega_n \mathrm{TE} + \Delta\phi_0 \right)\right]^2 + \left[fF\sum_{n=1}^{N} A_n \sin \left(\Delta \omega_n \mathrm{TE} + \Delta\phi_0\right) \right]^2 }
\label{eq:magphiFD2}
\end{equation}

\begin{equation}
    S_{m}(\mathrm{TE}, \mathrm{TR}, \theta)=S_{0} e^{-R_2^{*} \mathrm{TE}} \sqrt{\left[(1 - f)W + fF\sum_{n=1}^{N} A_n \cos \left(\Delta \omega_n \mathrm{TE}\right)\right]^2 + \left[fF\sum_{n=1}^{N} A_n \sin \left(\Delta \omega_n \mathrm{TE}\right) \right]^2}
    \label{eq:magFD2}
 \end{equation}

\begin{equation}
F(\mathrm{TR}, \theta)=\frac{1-e^{-R_{1f}\mathrm{TR}}}{1-e^{-R_{1f}\mathrm{TR}}\cos\theta} \sin\theta
\label{eq:F2}
 \end{equation}
\begin{equation}
W(\mathrm{TR}, \theta)=\frac{1-e^{-R_{1w}\mathrm{TR}}}{1-e^{-R_{1w}\mathrm{TR}}\cos\theta} \sin\theta
\label{eq:W2}
 \end{equation}

\subsection*{Phantom construction}

 Phantom experiments were carried out to assess the impact of the method used to provide initial parameter guesses, and to compare the three signal Fat DESPOT models, across a range of fat fractions. A phantom of fat-water emulsions was constructed following a published protocol
 \cite{bush_fat-water_2018} with slight adjustments. To facilitate the emulsification, the surfactant polyethylene glycol sorbitan monolaurate (Tween 20, MilliporeSigma Canada Ltd.) was added to the peanut oil. Likewise, the surfactant sorbitan monooleate (span 80, MilliporeSigma Canada Ltd), and the preservative sodium benzoate (MilliporeSigma Canada Ltd.), were added to the agar solution. Additionally, gadobutrol (Gadovist, Bayer Healthcare) was added as a relaxation agent ([Gd$^+$] = 0.2 mM) in the agar gel preparation. Each emulsion was placed in a 50 mL conical polypropylene tube (Corning\textregistered\ 50 mL centrifuge tubes). Two additional tubes, one containing pure peanut oil and one containing the agar gel preparation with gadobutrol and surfactants were also prepared, for a total of seven nominal fat volume fractions. The tubes were then suspended on a plastic and polystyrene rig placed in a cylindrical acrylic phantom container (Magphan \textregistered SMR170, The Phantom Laboratories, Salem, USA), which was then filled with a solution of distilled deionized water with gadobutrol ([Gd$^+$] = 0.3 mM) and sodium chloride (Windsor Salt Ltd) ([NaCl] = 24 mM) to approach the conductivity of human tissue \cite{duan_characterization_2014}. 

\subsection*{Phantom data acquisition}

All measurements of the phantom were performed at room temperature in a 3 T MRI scanner (Ingenia, Philips Healthcare) using a vendor-provided 15-channel receive-only head coil. The phantom was left to rest in the centre of the bore for at least 30 minutes before measurements to reduce flow artifacts. For Fat DESPOT measurements, a 3D mGRE sequence with monopolar readout and default gradient and RF spoiling was employed. Measurements were collected with four excitation FAs. Eight signal averages were acquired for each measurement, and parallel imaging was not used, to maximize SNR. Sequence parameters used in these experiments are summarized in Table \iftoggle{mrm}{S1}{\ref{s-tab:sequence_params}}.

A first data set was acquired to compare the impact of initial guesses provided by two fat-water separation approaches, 3-point Dixon and GC, when used with Fat DESPOT$_m$. Two 6-echo (2$\times$6) series were acquired at each FA using the a 3D mGRE with $\Delta$TE = 2.4 ms and TR = 18 ms. For the first acquisition, the initial echo time (TE$_1$) = 1.5 ms, and for the second acquisition, TE$_1$ = 2.7 ms. These TE$_1$s were selected such that the two acquisitions could be combined in post-processing to create a (2$\times$6) 12-echo train with shorter apparent $\Delta$TE (= 1.2ms). This short $\Delta$TE is required for a robust fat-water separation using 3-point Dixon \cite{fortier_mr-oximetry_2023}.

FAs were optimized to minimize $R_1$ estimate variance considering the TR and the expected range of $R_1$, for a set of four FAs \cite{deoni_rapid_2003, cheng_rapid_2006}. In phantoms, a range of 0.54--2.9 s$^{-1}$ was used based on other fat-water phantoms \cite{fortier_mr-oximetry_2023, franconi_tissue_2018}. Selected FAs were $\theta$=[3$^\circ$, 6$^\circ$, 15$^\circ$, 34$^\circ$]. 

In the second experiment, Fat DESPOT$_{m}$, Fat DESPOT$_{m\phi}$, and Fat DESPOT$_{c}$  were compared using data collected with an 8-echo 3D mGRE sequence previously designed by our group for a single acquisition\cite{ciobanu_evaluation_2022}. This acquisition scheme had four FAs, eight echoes, $\Delta$TE = 1.8 ms, TE$_1$ = 1.9 ms, and used the minimum TR = 24 ms.
FAs were reoptimised for TR=24 ms, resulting in angles $\theta$=[3$^\circ$, 7$^\circ$, 17$^\circ$, 39$^\circ$]. 

Experimental VFA $R_1$ measurements are known to be affected by $B_1$-induced flip angle variations, which were corrected throughout with a relative $B_1$ map from the dual-angle method\cite{samson_dam_b1_2006}. The $B_1$ map acquisition was done using a multi-slice turbo spin-echo (MS TSE) acquisition at two angles (FA = 60, 120$^\circ$), and other parameters noted in Table \iftoggle{mrm}{S1}{\ref{s-tab:sequence_params}}.

A separate series was acquired with a unipolar 3D mGRE sequence (TE$_1$ = 1 ms, \# of echoes = 6, $\Delta$TE=1.7 ms, FA=3$^\circ$) to provide a reference measurement for PDFF.

Phantom measurements had an acquired voxel size of 2.0$\times$2.0$\times$5.0 mm$^3$ and a reconstructed voxel size of  1.875$\times$1.875$\times$5.0 mm$^3$, covering a field of view (FOV) of 210$\times$210$\times$100 mm$^3$ for mGRE measurements and 210$\times$210$\times$90 mm$^3$ for MS TSE measurements.
The total scan times were 43.7 minutes for the 2$\times$6-echo datasets, and 30.78 minutes for the 8-echo datasets.
The additional scan time for the reference measurement for PDFF was 5.6 minutes. 

\subsection*{In vivo data acquisition}

Pilot measurements using Fat DESPOT$_c$ were conducted in the lower leg of a healthy volunteer (male, age 24) using the same 3 T MRI scanner as the phantom measurements and an 8-channel receive-only extremity coil. Vendor supplied uniformity correction was used (CLEAR, Phillips Healthcare). The same 8-echo acquisition protocol as the second phantom experiment was used, excluding the PDFF reference measurement. The study was approved by the Research Ethics Board of the McGill University Health Centre, and the volunteer gave informed consent. FAs were reoptimized to the expected $R_1$ range of human tissue, 0.56–3.33 s$^{-1}$ \cite{oconnor_comparison_2009, tadamura_effect_1997, ding_simultaneous_2013, bane2016feasibility, o2007organ, mcgrath2008oxygen, noseworthy1999tracking, little2018mapping, jones2002imaging, qian2020vivo, huen2014absence} with TR = 24 ms. This resulted in FAs of $\theta$=[4$^\circ$, 10$^\circ$, 22$^\circ$, 51$^\circ$] for Fat DESPOT$_c$. A smaller FOV (192.5$\times$160.4$\times$100 mm$^3$) was selected, reducing acquisition time. All other parameters were preserved from the acquisition protocol for phantom measurements, including the number of averages (8) and the absence of parallel imaging. The total scan time was 22.90 minutes, including $B_1$ mapping. 

\subsection*{Data processing}

The mGRE data underwent preprocessing steps before Fat DESPOT fitting. For the first phantom experiment, the two acquisitions taken at each FA were combined into a single data set for each FA by alternating echoes in increasing echo order. 

For all approaches, the relative $B_1$ map was constructed from the dual angle MS TSE acquisition and used to scale the nominal FA for each voxel \cite{boudreau_b1_2017}. For the in vivo experiment, all images were registered to the 3$^\circ$ mGRE Fat DESPOT acquisition using rigid registration (function \textit{imregtform} with default parameters, MATLAB 2023).  

Next, fat water separation was performed on the lowest FA data (=3$^\circ$) using either the 3-point Dixon \cite{berglund_three-point_2010} or GC algorithm
\cite{hernando_addressing_2012}. Maps of PDFF were calculated from the fat-water separated complex signal maps using a magnitude discrimination method \cite{liu_fat_2007}, to be used as an initial guess for Fat DESPOT. 

In all instances, maps of the apparent $R_1$ of the mixture, $R_{1global}$, were calculated using a VFA approach with data from the first echo \cite{wang2008rapid, fortier_mr-oximetry_2023}. The initial guesses for $R_{1f}$ and $R_{1w}$ for each voxel were set based on the previously generated estimate of PDFF. For PDFF$>$ 50\%, $R_{1global}$ was used as the guess for $R_{1f}$ and a fixed value of 1 s$^{-1}$ for $R_{1w}$. For PDFF$<$ 50\%, $R_{1global}$ was used as the guess for $R_{1w}$ with a fixed value of 4 s$^{-1}$ for $R_{1f}$. 

$R_2^*$ mapping for the initial guess was achieved with a monoexponential fit to data with TE$_1$=1.5 ms, FA=3$^\circ$ when the 3-point DIXON was used for the PDFF initial guess map. When GC was used to generate initial PDFF maps, the $R_2^*$ initial guess was obtained from the output of the GC algorithm.

$\phi_{0f}$ and $\phi_{0w}$ were also calculated from the 3$^\circ$-FA data to calculate $\Delta\phi_0$ when used for Fat DESPOT$_{m\phi}$ and to be used directly in Fat DESPOT$_c$. Finally, for Fat DESPOT$_c$, additional $B_0$ maps ($\Delta B_{0,\theta}$) were calculated from the GC output at each flip angle. 

Following this pre-processing, the Fat DESPOT$_m$, Fat DESPOT$_{m\phi}$, and Fat DESPOT$_c$ models were fit to their respective mGRE data using a non-linear least-squares algorithm (function \textit{lsqnonlin} using the \textit{trust-region-reflective} algorithm, MATLAB 2023). For phantom experiments, a six-resonance 
model of the chemical shift spectrum of peanut oil, with chemical shifts ($\delta_n= \Delta \omega_n / \omega_0$) = [0.80 ppm, 1.20 ppm, 2.00 ppm, 2.66 ppm, 4.21 ppm, 5.20 ppm,] and amplitudes ($A_n$) = [0.087, 0.694, 0.128, 0.004, 0.039, 0.048], was borrowed from previous experimental measurement \cite{triay_bagur_magnitude-intrinsic_2019}. For the in vivo measurement, a six-resonance fat spectrum from skeletal muscle, with $\delta_n$=[5.3 ppm, 4.13 ppm, 2.78 ppm. 2.24 ppm, 1.3 ppm, 0.9 ppm] and $A_n$=[0.066, 0.035, 0.011, 0.052, 0.077, 0.047, 0.598, 0.089], was used \cite{triplett_chemical_2014}. Setting upper and lower bounds for PDFF to be within 5\% of the PDFF initial guess was found to improve the accuracy of PDFF output compared to the reference measurement. All other parameter bounds are displayed in Table \iftoggle{mrm}{S2}{\ref{s-tab:bounds}}. In these experiments, the $B_0$ field map (obtained from fat-water separation) was observed to vary between flip angle acquisitions, and so the FA-specific $\Delta B_{0,\theta}$ maps were included in the fit as fixed parameters. The normalized root mean squared error (nRMSE) was used as to measure of fit quality. Normalization, relative to the magnitude of the signal from the first echo with the smallest flip angle, was used for ease of comparison between experiments. 

The reference PDFF measurement was obtained from the 6-echo unipolar mGRE data described above, using GC fat-water separation. 

Resulting multiparametric maps have been displayed using perceptually uniform colour maps\cite{green_crameri_2022, crameri_geodynamic_2018}.

\subsection*{Statistical analysis}

\begin{figure}[ht]
\begin{center}
 \includegraphics[width=\linewidth]{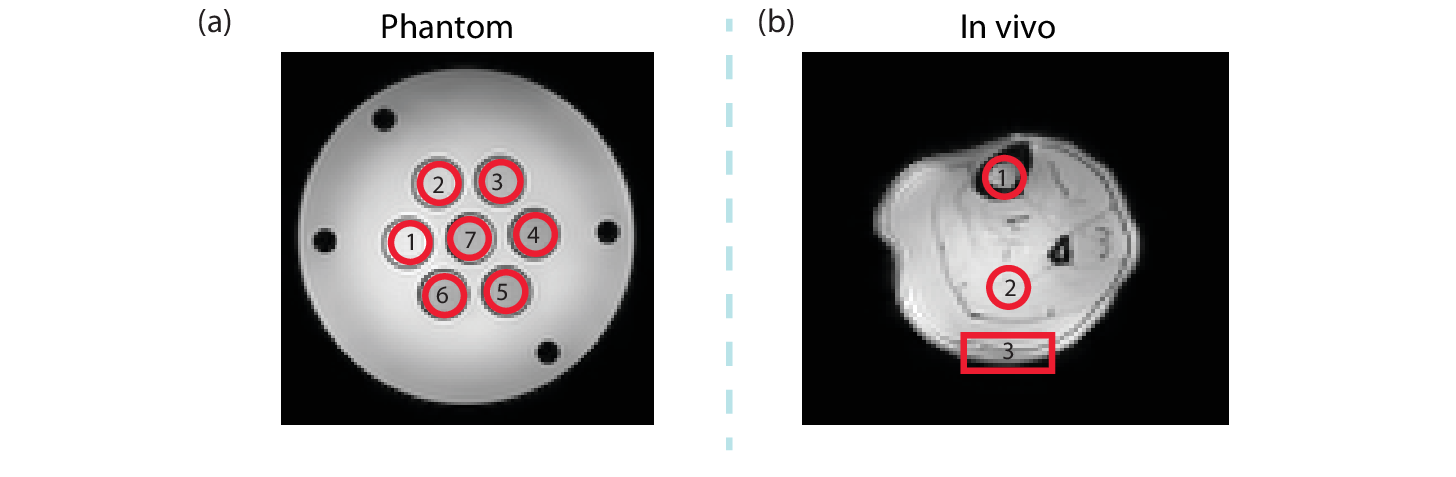}
 \end{center}
  \caption{Regions of interest of (a) the variable fat fraction phantom and (b) the lower leg of a human volunteer. In the phantom, ROIs 1-7 correspond to nominal fat fractions of 0\%, 5\%, 25\%, 50\%, 60\%, 75\%, and 100\% respectively. In the lower leg, ROIs 1-3 correspond to tubular bone marrow, calf skeletal muscle, and subcutaneous fat. Bone marrow and subcutaneous fat voxels of interest within the ROI were selected based on a PDFF estimate $>$70\% and $>60$ respectively. All ROIs were measured over 3 slices of the acquired image.}
  \label{fig:ROIs}
\end{figure}

For quantitative measurements and statistical analysis of PDFF, $R_2^*$, $R_{1f}$, and $R_{1w}$, regions of interest (ROIs) were selected (Figure \ref{fig:ROIs}). In the phantom, manually drawn circular ROIs with matched volumes (number voxels = 243) were selected to fit within the cross-sectional area of each tube. In vivo, circular ROIs were selected for the bone marrow (in the tibia) and the soleus muscle, and a rectangular ROI for the subcutaneous fat. Due to the size and shape of the bone marrow and subcutaneous tissue, a geometrical ROI could not be used alone without significantly reducing the number of voxels included in the quantitative analysis or including voxels from other tissues. Hence, semi-automatic ROIs were created by including voxels within the geometrical shapes with PDFF $>$ 60\% for the subcutaneous fat and PDFF $>$ 70\% for the bone marrow. These thresholds were selected to include the full range of PDFFs expected based on the subcutaneous fat (74$\pm$13\% \cite{de_bazelaire_mr_2004}) and bone marrow (between 82\% and 94\% \cite{held_intraindividual_2022, franz_association_2018}) reported in the literature. The thresholds were also cross-referenced with the observed distribution in a PDFF value histogram of the larger selection areas. This ensured that the voxels included in the analysis were representative of the tissue of interest. All ROIs were measured over three slices selected centrally to the imaging volume. 

The mean and standard deviation of each parameter, and the nRMSE were calculated for each ROI in phantoms and in vivo. The combined nRMSE within each ROI was obtained using equation \ref{eq:nRMSE_comb} where $K$ is the number of nRMSEs being combined. Statistical analysis was carried out for experiments in phantoms. Comparison of means was conducted for each parameter in the model using a two-way ANOVA (function \textit{anova2}, MATLAB 2023) to assess the effects of Fat DESPOT approach, ROI (nominal fat fraction), and interactions. Inter-technique means were then compared using multiple pair-wise testing with Bonferroni correction (function \textit{multcompare} with \textit{CriticalValueType bonferroni}, and \textit{anova1} input MATLAB 2023) and a $p$-value of 0.05 was used to determine significance. To compare standard deviations between approaches, a two-sample F-test for equal variance (function \textit{vartest2} MATLAB 2023) was used with a significance threshold of 0.05. To compare variability across models, the combined standard deviation across ROIs was calculated for each approach following equation \ref{eq:sum_sd}, where $N$ is the number of ROIs being combined, $w$ is the sample size of the ROI, and $\sigma$ is the standard deviation.

\begin{equation}
    nRMSE_{combined}=\sqrt{\frac{ \sum_{k=1}^K nRMSE_k^2}{K}}
    \label{eq:nRMSE_comb}
\end{equation}

\begin{equation}
    m_{combined}=\sqrt{\frac{\sum_{n=1}^N w_n \sigma_n^2}{\sum_{n=1}^N w_n}}
    \label{eq:sum_sd}
\end{equation}

\section*{Results}
\subsection*{Comparing initial estimates from 3-point Dixon and GC}

\begin{figure}[hp]
    \begin{center}
     \includegraphics[width=\linewidth]{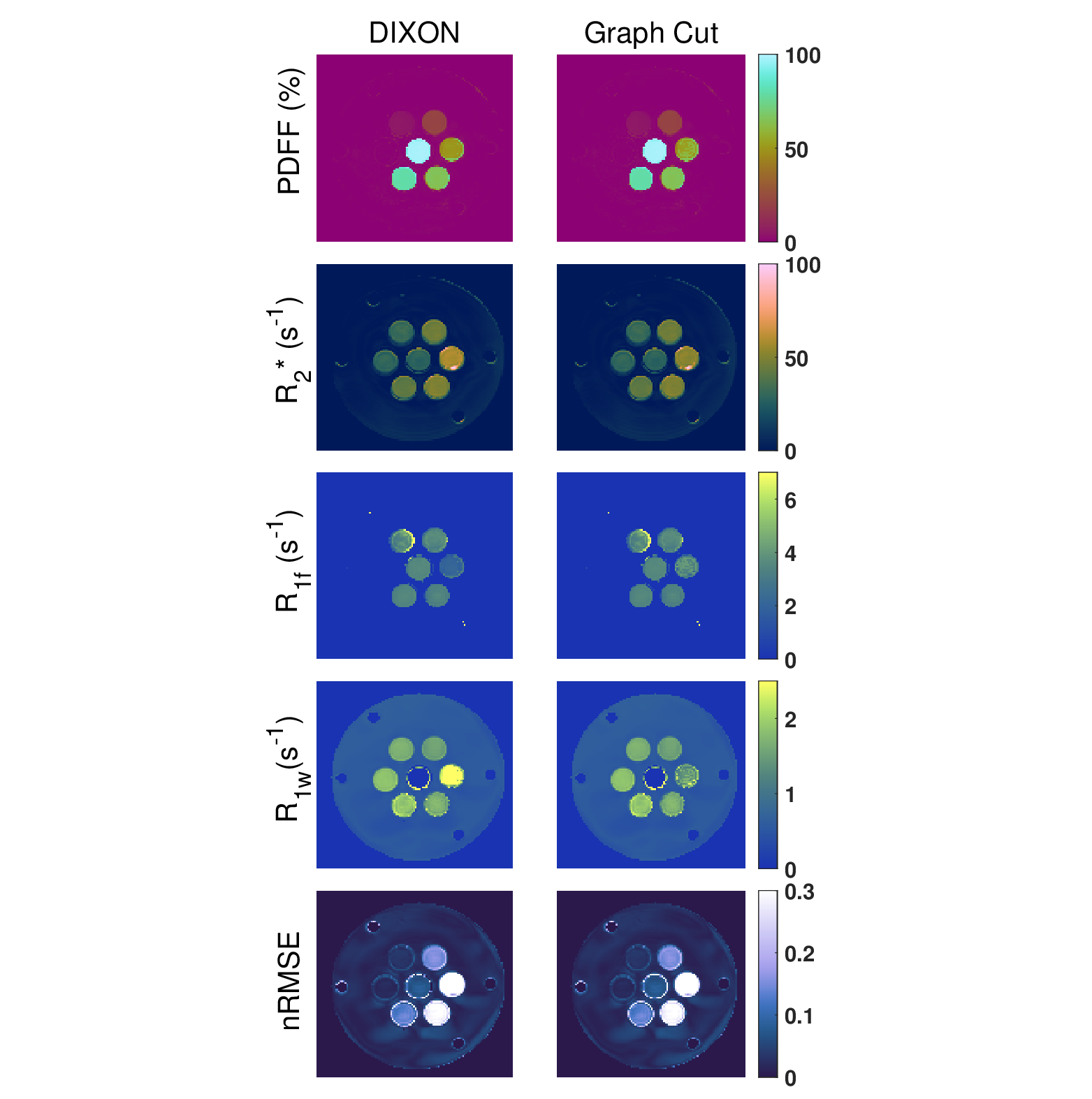}
     \end{center}
      \caption{Multiparametric maps for PDFF, $R_2^*$, $R_{1f}$, $R_{1w}$, and nRMSE using the 3-point Dixon and GC as PDFF initial guess inputs for Fat DESPOT$_m$ on a 2$\times$6-echo dataset. To reduce noise in the $R_{1f}$ images, voxels with PDFF$<$3\% and in the $R_{1w}$ images, voxels PDFF$>$97\% were masked.
      }
      \label{fig:phantom_maps_DIXON}
\end{figure}

Analysis of the 2$\times$6-echo data analyzed with the Fat DESPOT$_m$ model returned plausible parameters, regardless of the method used to generate initial guesses for PDFF (3-point Dixon or GC). This is reflected in the fairly uniform multiparametric maps (Figure \ref{fig:phantom_maps_DIXON}). Fit quality was similar across all ROIs, with a combined nRMSE of 0.20 using both approaches. The highest nRMSE was 0.38 using the DIXON approach and 0.37 using the GC approach, observed in the ROI with a nominal fat fraction of 50\%. This tube also had the highest difference in estimates across all measured parameters.  In the $R_2^*$ maps, artifacts appeared in the water compartment surrounding the emulsion tubes, likely due to $B_0$ field inhomogeneity from the styrofoam support in the phantom not sufficiently accounted for by the magnitude model fit, regardless of the choice of Dixon or GC. 

\begin{figure}[hp]
    \begin{center}
     \includegraphics[width=\linewidth]{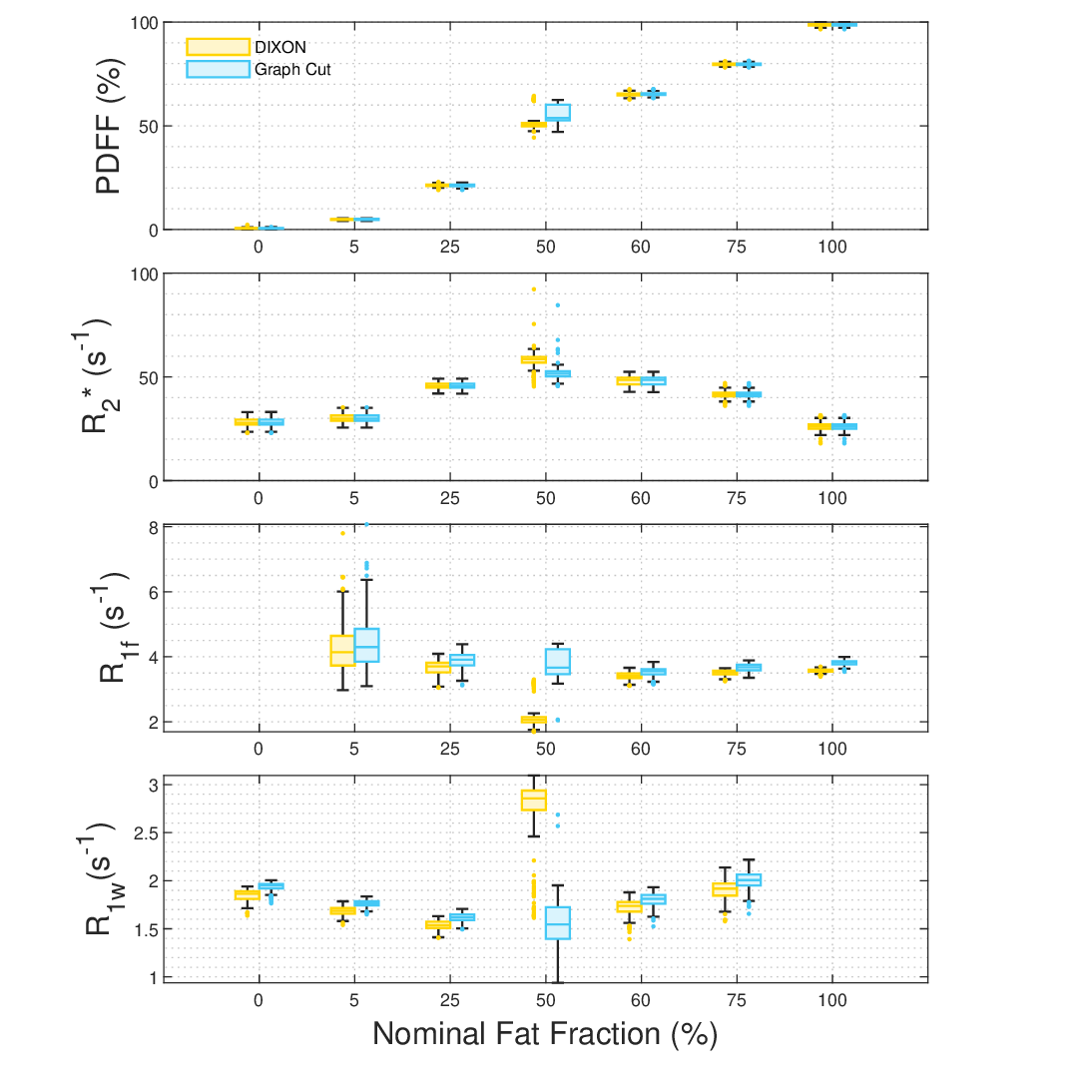}
     \end{center}
      \caption{Distribution of voxel-wise estimates of PDFF, $R_2^*$, $R_{1f}$, and $R_{1w}$, using the 3-point Dixon and GC as PDFF initial guess inputs for Fat DESPOT$_m$ on a 2$\times$6-echo dataset.  Box = interquartile range, horizontal line = median, feathers= 1st and 4th quartile, dots= outliers.}
      \label{fig:phantom_box_DIXON}
\end{figure}

Values of the estimated parameters from the the voxel-wise Fat DESPOT$_m$ fit are displayed in Figure \ref{fig:phantom_box_DIXON}, extracted from ROIs shown in Figure \ref{fig:ROIs}.a. The choice of Dixon or GC fat-water separation to generate initial guesses of PDFF for Fat DESPOT$_m$ returned relatively stable and similar estimates of PDFF, $R^*_2$, $R_{1f}$ and $R_{1w}$ across fat fractions. Average percent differences were 11.6$\pm$1.3\% for $R_{1f}$ and 11.9$\pm$2.3\% for $R_{1w}$, respectively. The most obvious impact was in the ROI with nominal fat fraction = 50\%, where values calculated using the Dixon approach deviated considerably from the overall trend. 

Two-way ANOVA conducted for each parameter with effects 'ROI', 'Approach' (Dixon or GC for initial guesses), and interactions, revealed a statistically significant effect of fitting approach for every parameter ($p<0.001$). Post-hoc testing revealed that the statistically significant effect of fitting approach on $R_{1f}$ and $R_{1w}$ was present across all but one ROI each. Conversely, PDFF and $R_2^*$ estimates were only significantly different in the ROI with 50\% nominal fat fraction ($p << 0$). When compared to the reference  measurement, PDFF from Fat DESPOT$_m$ was lower, with differences of -4.3$\pm$4.0 (initial guess from 3-point Dixon) and -1.0$\pm$4.0 (initial guess from GC), both statistically significant.

Precision was similar between techniques. Standard deviations for PDFF were not significantly different, and significantly higher for $R_2^*$ (from GC initial guess) only in the 50\% nominal fat fraction ROI ($p=2\times 10^{-3}$). $R_{1f}$ from vials with nominal fat fractions of 25\%, 50\%, 75\%, and 100\% using GC initial guesses had significantly higher standard deviations than with DIXON (0.1 s$^{-1}$ vs. 0.11 s$^{-1}$ on average; $p\leq0.03$). Conversely, the GC approach $R_{1w}$ returned lower standard deviations in vials with nominal fat fractions of 0\%, 5\%, and 50\% (0.033 s$^{-1}$ vs. 0.043 s$^{-1}$ on average; $p\leq10^{-3}$ ).

\subsection*{Comparing Fat DESPOT$_m$, Fat DESPOT$_{m\phi}$ and Fat DESPOT$_c$}

Inspection of the outputs of GC fat-water separation revealed that the initial phase of fat and water signals were drastically different. This can be seen in examples of initial guess maps from GC and DESPOT1 
(Figure \iftoggle{mrm}{S1}{\ref{s-fig:initialguess_phantom}}). 
This suggests that initial phase should be a distinct free parameter in the Fat DESPOT$_c$ model.

\begin{figure}[hp]
    \begin{center}
     \includegraphics[width=\linewidth]{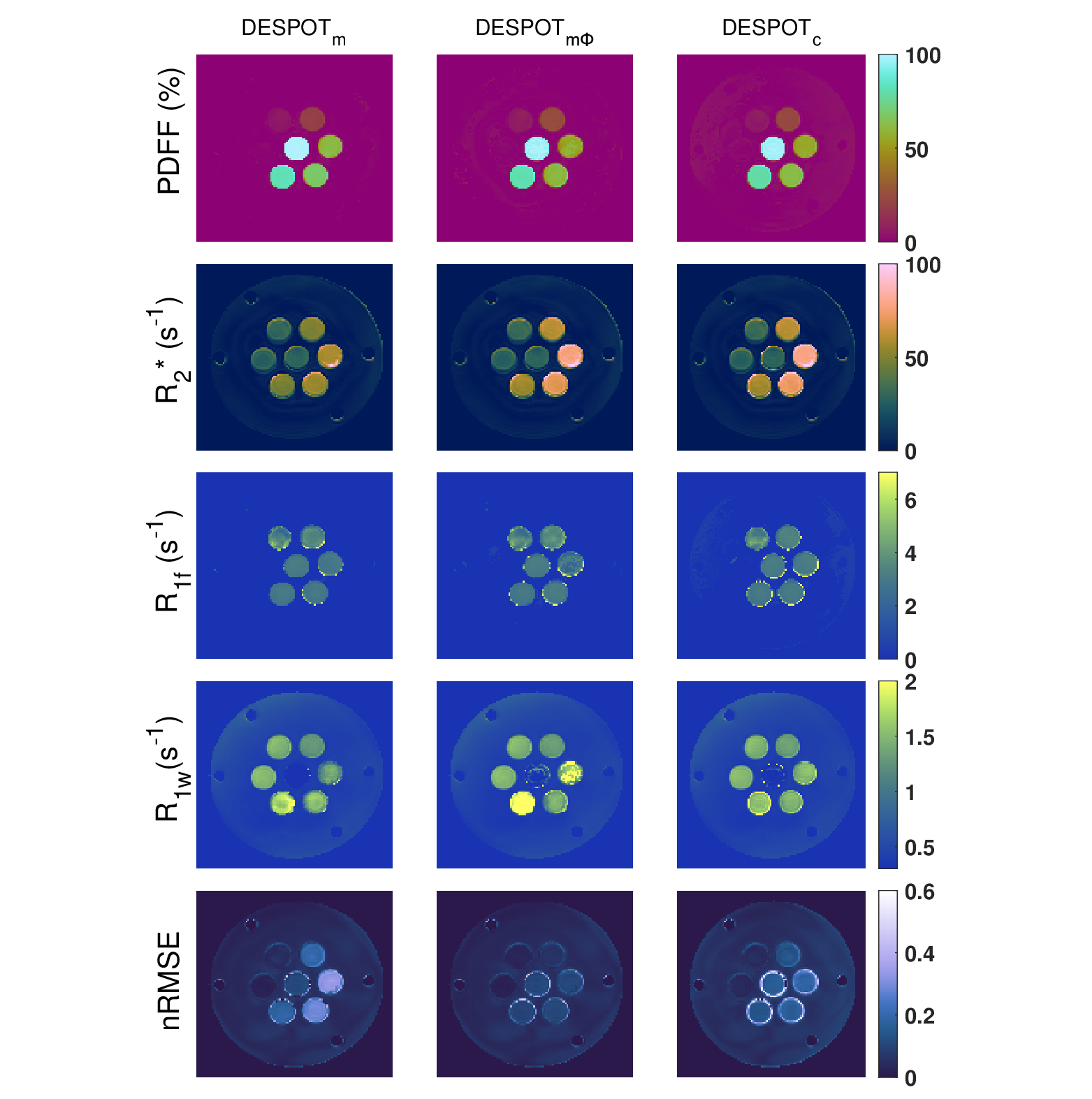}
     \end{center}
      \caption{Multiparametric maps for PDFF, $R_2^*$, $R_{1f}$, $R_{1w}$, and nRMSE using Fat DESPOT$_m$, Fat DESPOT$_{m\phi}$, and Fat DESPOT$_c$ on an 8-echo dataset. To reduce noise in the $R_{1f}$ images, voxels with PDFF $<$ 3\% and in the $R_{1w}$ images, voxels PDFF $>$ 97\% were masked. Voxels outside the phantom were masked.}
      \label{fig:phantom_maps}
\end{figure}

Maps of PDFF, $R_2^*$, $R_{1f}$, and $R_{1w}$ obtained from Fat DESPOT$_m$, Fat DESPOT$_{m\phi}$, and Fat DESPOT$_c$ fits to the 8-echo acquisitions (Figure \ref{fig:phantom_maps}) demonstrated high-quality fits in all ROIs. This being said, Fat DESPOT$_{m\phi}$ and Fat DESPOT$_c$, which consider phase differences between fat and water, had a lower combined nRMSE, 0.08 and 0.14 respectively, compared to Fat DESPOT$_m$ with an average value of 0.20. The highest average nRMSE for a single ROI was 0.33 for Fat DESPOT$_m$, 0.11 for DESPOT$_{m\phi}$, and 0.20 for Fat DESPOT$_c$, corresponding to the tube with a nominal fat fraction of 50\%. As with the previously discussed 2$\times$6-echo data fitted with the Fat DESPOT$_m$ approach, artifacts appeared in the water compartment of the $R_2^*$ map regardless of the Fat DESPOT approach. Example model fits using Fat DESPOT$_m$, Fat DESPOT$_{m\phi}$, and Fat DESPOT$_c$ in single voxels of 3 ROIs are plotted in Figure \iftoggle{mrm}{S2}{\ref{s-fig:voxelwise}}. 

\begin{figure}[hp]
    \begin{center}
     \includegraphics[width=\linewidth]{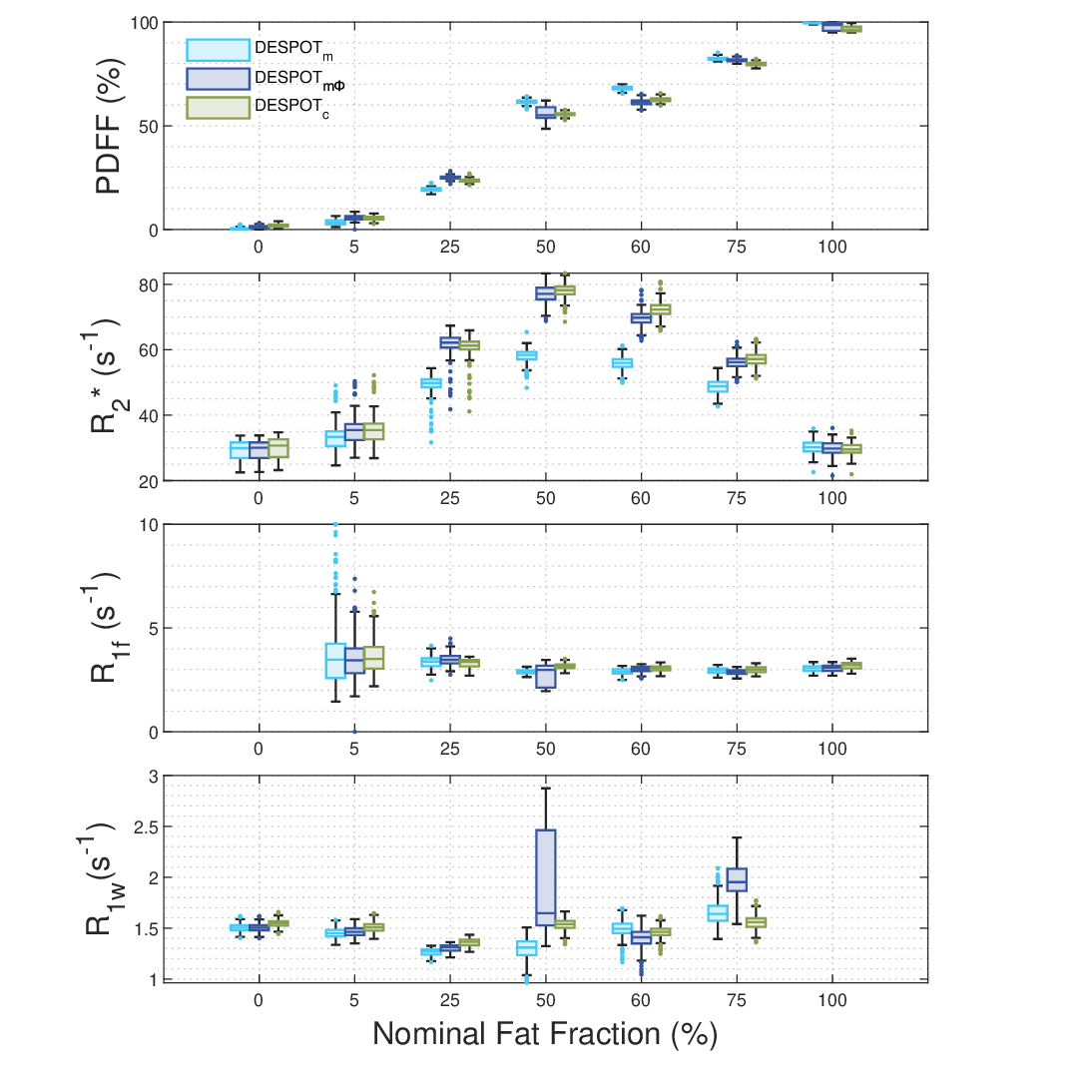}
     \end{center}
      \caption{Distribution of voxel-wise estimates of PDFF, $R_2^*$, $R_{1f}$, and $R_{1w}$, using Fat DESPOT$_m$, Fat DESPOT$_{m\phi}$, and Fat DESPOT$_c$ on an 8-echo dataset.  Box = interquartile range, horizontal line = median, feathers= data range, dots= outliers.}
      \label{fig:phantom_box}
\end{figure}

Parameter estimates from fits to the 8-echo data showed similar trends versus fat fraction as those from the 12-echo data. Parameter values are displayed in Figure \ref{fig:phantom_box}. The most obvious deviations between Fat DESPOT model outputs were observed for PDFF and $R^*_2$ from Fat DESPOT$_m$ (in ROIs with 25--75\% nominal fat fraction) and for $R_{1,f}$ and $R_{1,w}$ in select ROIs. Two-way ANOVA (factors 'ROI' and 'Model') for each parameter revealed statistically significant effects for all parameters ($p \leq 10^{-18}$). Post-hoc testing confirmed that differences between all three pairs of approaches, for all parameters, were statistically significant. Despite this, PDFF, $R_2^*$, values from phase-sensitive methods Fat DESPOT$_{m\phi}$ and Fat DESPOT$_c$ were far more similar to each other than to those from Fat DESPOT$_m$ (Table \ref{tab:reldiff}). The only exception was $R_{1f}$, where Fat DESPOT$_m$ and Fat DESPOT$_c$ were more similar to each other than to Fat DESPOT$_{m\phi}$. 

\begin{table}[ht]

\caption{Relative difference in estimates of PDFF, $R_2^*$, $R_{1f}$, and $R_{1w}$ between pairs of approaches}
\label{tab:reldiff}
\begin{tabular}{lrrr}
\hline
          & \multicolumn{3}{c}{Fat DESPOT approach pairings (\%)} \\ \hline
parameter & m, m$\phi$            & m, c                 & m$\phi$, c           \\ \hline
PDFF (\%)      & 30.8             & 33.0            & 7.9              \\
$R_2^*$ (\%)   & 13.4            & 14.6           & 1.6            \\
$R_{1f}$ (\%)  & 13.9            & 6.0            & 13.0           \\
$R_{1w}$ (\%)  & 34.7             & 26.0            & 20.8            \\ \hline
\end{tabular}
\end{table}

\begin{figure}[htb]
    \begin{center}
     \includegraphics[width=\linewidth]{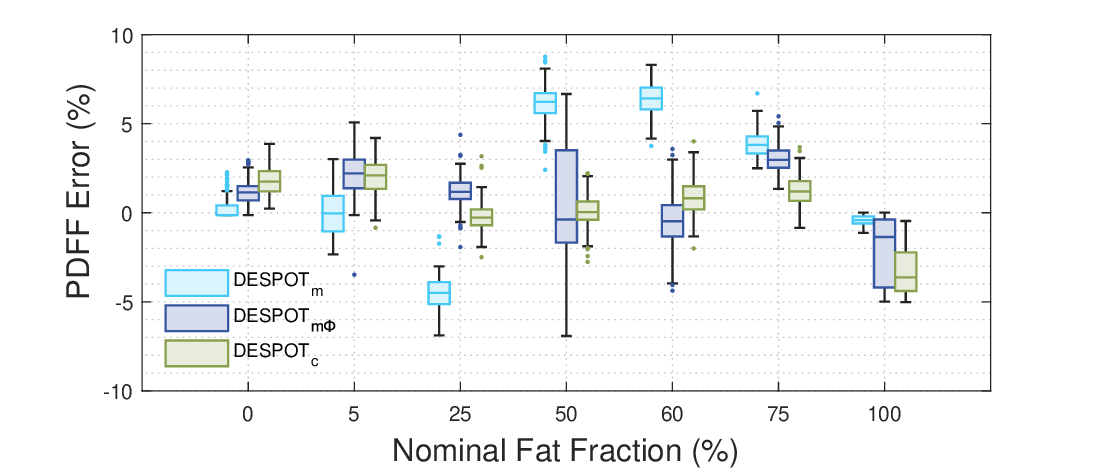}
     \end{center}
     \caption{Distribution of the error on the PDFF using Fat DESPOT$_m$, Fat DESPOT$_{m\phi}$, and Fat DESPOT$_c$ on an 8-echo dataset compared to a reference measurement. Box = interquartile range, horizontal line = median, feathers= data range, dots= outliers. Error is significantly different (p$<$0.05) between approaches in all ROIs excluding the 75\% nominal fat fraction.}
      \label{fig:phantom_pdff}
\end{figure}

PDFF estimates were significantly different from the reference measurement for all Fat DESPOT models ($p=0$ for all approaches) (Figure \ref{fig:phantom_pdff}). Fat DESPOT$_m$ exhibited the highest mean error of 3.2$\pm$2.5 compared to 1.9$\pm$1.4\% for Fat DESPOT$_{m\phi}$ and 1.5$\pm$1.2\% for Fat DESPOT$_c$.

Overall, Fat DESPOT$_c$ resulted in lower standard deviation for PDFF, $R_{1f}$, and $R_{1w}$. Fat DESPOT$_m$ had the lowest standard deviation for $R_2^*$, and Fat DESPOT$_{m\phi}$ had the highest standard deviation for all parameters (Table \ref{tab:phantom_std}). However, not all ROIs had statistically significant differences in standard deviation and $R_2^*$ standard deviation differences were only significantly different for 1/7 ROIs when comparing magnitude approaches to the complex approach and 2/7 ROIs when comparing magnitude approaches to each other. Fat DESPOT$_{m\phi}$ has higher variability in standard deviations across ROIs, suggesting lower fit stability. This is particularly noticeable in the precision of $R_{1w}$ estimates for the 50\% and 75\% tubes. 

\begin{table}[h]
\caption{The combined standard deviations for PDFF, $R_2^*$, $R_{1f}$, and $R_{1w}$ across ROIs 1-7 in the variable fat fraction phantom for Fat DESPOT$_m$, Fat DESPOT$_{m\phi}$, and Fat DESPOT$_c$ }
\label{tab:phantom_std}
\begin{tabular}{lrrr}
\hline
parameter          & Fat DESPOT$_m$ & Fat DESPOT$_{m\phi}$ & Fat DESPOT$_c$ \\ \hline
PDFF (\%)          & 0.14           & 0.24                 & 0.13           \\
$R_2^*$ (s$^{-1}$) & 0.44           & 0.47                 & 0.47           \\
$R_{1f}$ (s$^{-1}$)          & 0.28           & 0.21                 & 0.19           \\
$R_{1w}$ (s$^{-1}$)           & 0.013          & 0.0340               & 0.0082         \\ \hline
\end{tabular}
\end{table}

\subsection*{In vivo results}

\begin{figure}[hp]
    \begin{center}
     \includegraphics[width=\linewidth]{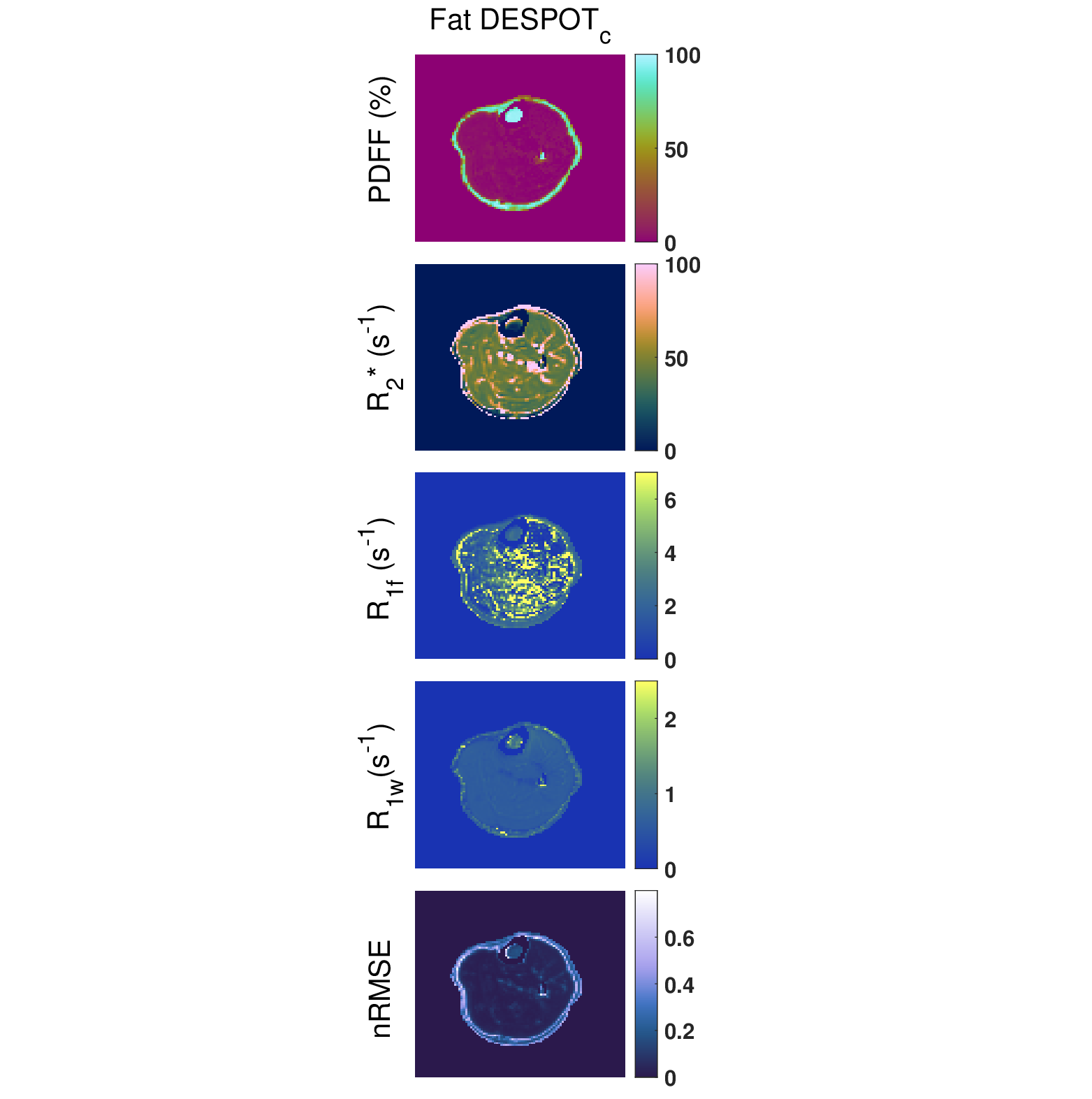}
     \end{center}
      \caption{Multiparametric maps of a cross-section of the lower leg for PDFF, $R_{2}$*, $R_{1f}$, $R_{1w}$, and nRMSE using Fat DESPOT$_c$. Voxels outside the leg were masked.}
      \label{fig:invivo_maps}
\end{figure}

In vivo Fat DESPOT$_c$ multiparametric maps of the lower leg (Figure \ref{fig:invivo_maps}) displayed key anatomical features, including the muscle, bone marrow from the tibia, the fibula, and the subcutaneous fat layer with distinct combinations of Fat DESPOT output values. Initial guess maps are presented in Figure \iftoggle{mrm}{S3}{\ref{s-fig:initialguess_invivo}}. The mean value and standard deviation of Fat DESPOT output parameters are displayed in Table \ref{tab:invivo_dispersion}. 

\begin{table}[ht]
    \caption{Mean value of Fat DESPOT$_c$ output parameters PDFF, $R_{2}$*, $R_{1f}$, and $R_{1w}$ and mean nRMSE for ROIs in the bone marrow, muscle, and subcutaneous fat of a human lower leg.}
    \label{tab:invivo_dispersion}
\begin{tabular}{lrrr}
\hline
\multicolumn{1}{c}{Parameter} & \multicolumn{3}{c}{ROI}                                                                             \\ \hline
                              & \multicolumn{1}{c}{Bone Marrow} & \multicolumn{1}{c}{Muscle} & \multicolumn{1}{c}{Subcutaneous Fat} \\ \cline{2-4} 
\# Voxels                     & 143                             & 243                        & 168\\
PDFF (\%)                     & 95$\pm$4& 1.9$\pm$1.3& 88$\pm$9\\
$R_2^*$ (s$^{-1}$)            & 40$\pm$50& 46$\pm$10& 17$\pm$10\\
$R_{1f}$(s$^{-1}$)            & 2.0$\pm$0.6& 5$\pm$3& 2.3$\pm$0.4\\
$R_{1w}$(s$^{-1}$)             & 1.4$\pm$1.7& 0.60$\pm$0.03& 1$\pm$1\\
nRMSE& 0.26& 0.035& 0.24\\ \hline
\end{tabular}
\end{table}

\section*{Discussion}

This study sought to advance signal modeling for multiparametric mapping in fat-water mixtures. Two new signal models were introduced, notably to account for the initial phase of the signal components: one phase-sensitive magnitude model, and the other a fully complex model.

Fat DESPOT$_m$ with GC performed as well or better than with the 3-point Dixon. The 3-point DIXON requires specific echo-selection for best performance \cite{berglund_three-point_2010} which required a more time consuming 2$\times$6 echo acquisition, while GC does not and therefore is a superior choice for initial parameter estimation. Using the Dixon and GC algorithms to provide initial guesses for Fat DESPOT$_m$ returned similar results with parametric estimates agreeing on PDFF and $R_2^*$ values for nearly all ROIs and having low percent differences for $R_{1f}$ and $R_{1w}$ for almost all ROIs. However, the GC and 3-point Dixon approaches did not agree across any parameter in the 50\% tube. This suggests that inaccurate initial guesses of PDFF (from 3-point Dixon) or $R_2^*$ (from a monoexponential fit) may have affected $R_{1f}$ and $R_{1w}$ estimates in this tube. Furthermore, while precision was similar, GC performed slightly better overall in 2 of 3 parameters, where standard deviations were significantly lower.
Finally, our GC protocol uses a single 8-echo acquisition scheme rather than the 2$\times$6-echo proposed in previous work\cite{fortier_mr-oximetry_2023}. Reducing the number of acquisitions required for Fat DESPOT increases the clinical feasibility of the approach.

The Fat DESPOT$_c$ approach, which obtains an initial guess of PDFF, $R_2^{*}$, $B_{0\theta}$, $\phi_{0f}$, and $\phi_{0w}$ from GC and fits to the complex data, returns more accurate estimates of PDFF, and more precise estimates of $R_{1f}$ and $R_{1w}$ in phantoms. The resulting parametric estimates using Fat DESPOT$_{m\phi}$ had a larger standard deviation in some ROIs suggesting they were less stable compared to Fat DESPOT$_c$, or that the fully complex fit benefits from having more data points. 

The inclusion of phase in the Fat DESPOT model may also be responsible for some of the larger disagreements between approaches. Phase-sensitive models (Fat DESPOT$_{m\phi}$ and Fat DESPOT$_c$) returned higher $R_2^*$ compared to Fat DESPOT$_m$ in intermediate nominal fat fraction tubes, while no discernible trend in differences was noted for other parameters. This bias in $R_2^*$, paired with visible differences in the initial phase of fat and water from the GC initial guess suggests that error due to the inaccurate modeling of phase may be largely absorbed in the $R_2^*$ parameter. Furthermore, accounting for phase differences between fat and water resulted in higher quality fits when using both the complex and magnitude data, with Fat DESPOT$_{m\phi}$ having the highest fit quality among the approaches tested. Conversely, the lower precision of Fat DESPOT$_{m\phi}$ compared to all other techniques may be due to insufficient data to compensate for the inclusion of an additional parameter for phase in the model leading to overfitting when using the magnitude signal only.

Comparison of the relaxation parameter values measured in phantom in this work with the literature is complicated by contradictory trends in prior publications, and few reports of $R_{1w}$ and $R_{1f}$.  Indeed, some studies found that $R_{1w}$ in gel phantoms was independent of fat fraction \cite{roberts_confounder-corrected_2023, garrison_water-only_2022}, a behaviour consistent with our observations. Others found that both $R_{1f}$ and $R_{1w}$ were fat fraction-dependent in phantoms\cite{hu_change_2010}. However, the measurement approach and phantom construction, including the use of agar or agarose, dissolved contrast agents, and their respective concentrations, varied between studies, making comparison difficult. 

Comparing in vivo measurements to published data was challenging due to limited literature measuring PDFF, $R_2^*$, $R_{1f}$, and $R_{1w}$ in a single anatomical site, and due to potential inter-subject variations. This said, our measurement of $R_{1f}$ in tibia bone marrow agreed with a report of bone marrow $R_{1f}$ in the femur (3.8$\pm$1.3 s$^{-1}$ compared to 3.9$\pm$0.3 s$^{-1}$ in literature \cite{le_ster_breath-hold_2016}). Our $R_{1f}$ in subcutaneous fat (2.3$\pm$0.5s$^{-1}$) agreed with $R_{1global}$ estimates---which should be dominated by the fat signal---from one study (2.59 s$^{-1}$\cite{oconnor_comparison_2009}) but not with another (4.24 s$^{-1}$ \cite{tadamura_effect_1997}). Our $R_{1w}$ estimates in bone marrow were also similar to published values (1.4$\pm$1.7 s$^{-1}$ compared to 1.43$\pm$0.77 s$^{-1}$ \cite{le_ster_breath-hold_2016}), though the uncertainty on $R_{1w}$ in low-water-content environments is too large draw a strong conclusion. In muscle, where water content is higher, $R_{1w}$ from Fat DESPOT$_c$ was close to $R_{1global}$ in muscle---which should be dominated by the water signal--- from one source (0.59$\pm$0.03 s$^{-1}$ compared to 0.7 s$^{-1}$ \cite{gold_musculoskeletal_2004}) but less so for another source (1.13 s$^{-1}$ \cite{ding_simultaneous_2013}). Anecdotally, muscle is also the tissue ROI which showed the best overall fit quality.

In the bone marrow and subcutaneous fat,  $R_2^*$ was also similar to published values (47$\pm$50 $s^{-1}$ and 25$\pm$14 s$^{-1}$ compared to  60.8$\pm$5.1 s$^{-1}$\cite{meloni_quantitative_2022} and 23 s$^{-1}$ \cite{yu_multi-echo_2008} in literature for bone marrow and subcutaneous fat, respectively), though uncertainty was once again very high in the bone marrow. In muscle, our estimates were much higher than published values (46$\pm$10 s$^{-1}$ compared to literature reports of 24 s$^{-1}$\cite{zaeske_behaviour_2022} and 25 s$^{-1}$, \cite{ding_simultaneous_2013}). Discrepancies in bone marrow measurements may be due to the relatively small volume of bone marrow in the tibia leading to some averaging effects due to contamination from nearby tissues.

There are some limitations to the Fat DESPOT$_c$ approach presented here. In our experiments, the fat spectra were not measured directly, which may have affected the accuracy of the PDFF \cite{karampinos2018quantitative}. Furthermore, our model uses a single initial phase for all FAs. However, the initial phase is FA-dependent and could therefore be different in each acquisition \cite{wang_t1_2020}. When developing our model, we found that the inclusion of an initial phase for each FA resulted in unstable parametric estimates, but this simplification may have affected the quality of the fit. Additionally, all approaches to Fat DESPOT appear vulnerable to $B_0$ field inhomogeneity artifacts, notably from the styrofoam insert used in the phantom design. To reduce these issues, alternative initial parameter estimation techniques could be explored. Acquisition time remains a disadvantage in this implementation of Fat DESPOT; however, all acquisitions in this work used 8 signal averages and no parallel imaging, resulting in very high SNR. Assessment with shorter scan times (and lower SNR) is warranted. Fat DESPOT$_m$ has been found to perform well at an SNR above 63 \cite{fortier_mr-oximetry_2023}, and data acquisition in this work is equivalent to that prior work, such that there is reason to believe that performance can be maintained with shorter scans. Reduction of the number of averages and/or introduction of parallel imaging while keeping above this SNR threshold will allow for gains in the acquisition time without reduced fit quality. Furthermore, the number of FAs acquired and used in the fitting algorithm could be reduced \cite{fortier_mr-oximetry_2023}. Finally, while the lower leg provided a straightforward site for in vivo measurement, further experiments should be conducted in sites with a broader diversity of tissues, such as the abdomen, where the liver is of particular interest, given the emerging role of multiparametric mapping in the diagnosis of liver disease \cite{schaapman_multiparametric_2021, banerjee_multiparametric_2014}. This potential application will require careful consideration of motion issues and scan time.

\section*{Conclusion} 

Phase-sensitive modeling for Fat DESPOT, in particular the complex approach, offers higher precision and accuracy for phantom measurements compared to other versions of this technique and in vivo parametric estimates were comparable to literature. The 3D mGRE sequence used for Fat DESPOT is accessible on all clinical scanners, making it highly translatable. The use of the GC algorithm to calculate initial parameter guesses increases the flexibility of echo time selection, further simplifying data acquisition for this technique. Hence, the complex approach to Fat DESPOT represents a valuable advancement for multiparametric mapping with potential applications in fatty liver disease, and solid tumour imaging, where measures of $R_2$*, PDFF, $R_{1w}$, and $R_{1f}$ are of particular value.


\section*{Acknowledgments}

The authors acknowledge the developers of the ISMRM fat-water toolbox (http://www.ismrm.org/workshops /FatWater12/data.htm), Norma Ybarra for technical assistance in phantom building, and the MR Methods Research Group (McGill University) for useful discussion. This work was funded by the Research Institute of the McGill University Health Centre, the \textit{Fond de Recherche Québec - Santé} (FRQS), and a Discovery Grant from the Natural Sciences and Engineering Research Council of Canada (NSERC). RC Bider acknowledges support from NSERC (CGS-M award).


\newpage
	{
	
	\bibliography{references}
	\bibliographystyle{unsrt}
	
	}

\iftoggle{mrm}{
  \newpage
\renewcommand{\thepage}{S-\arabic{page}}
\setcounter{page}{1}

\section*{Supplementary Information}

\setcounter{figure}{0}
\renewcommand{\figurename}{Figure}
\renewcommand{\thefigure}{S\arabic{figure}}

\setcounter{table}{0}
\renewcommand{\tablename}{Table}
\renewcommand{\thetable}{S\arabic{table}}

Supplementary information for \textit{\mytitle}.

 \begin{table}[H]
\caption{Sequence parameters for complex and magnitude Fat DESPOT, B1 mapping, and unipolar FW separation in phantom and in vivo.}
    \centering
\begin{tabular}{lrrrr}
\hline
& 2$\times$6-echo& 8-echo& $B_1$ mapping                & \begin{tabular}[c]{@{}l@{}}Unipolar \\ FW separation\end{tabular}       \\ \hline
Acquisition type                                                                                           & mGRE & mGRE& MS TSE                       & mGRE                          \\
TR (ms)                                                                                                    & 18                            & 24                            & 1000                         & 18                            \\
TE$_1$ (ms)                                                                                                & 1.5, 2.7                      & 1.9                           & 9                            & 1.1                           \\
$\Delta$TE                                                                                                 & 2.4                           & 1.8                           & --                           & 1.7                           \\
\# TE                                                                                                      & 6$\times$ 2                   & 8                             & 1                            & 6                             \\
 NSA& 8& 8& 1&8\\
\begin{tabular}[c]{@{}l@{}}FA - Phantom \\ ($^\circ$)\end{tabular}                           & 3, 6, 15, 34                  & 3, 7, 17, 39                  & 60, 120                      & 3                             \\
\begin{tabular}[c]{@{}l@{}}FA - in vivo\\  ($^\circ$)\end{tabular} & 3, 8, 19, 45                  & 4, 10, 22, 51                 & 60, 120                      & 3                             \\
BW(Hz/px)                                                                                                & 1360                          & 1360                          & 1360                         & 1360                          \\
\begin{tabular}[c]{@{}l@{}}Recon. Voxel Size\\ - Phantom (mm$^3$)\end{tabular}& 1.875$\times$1.875$\times$5   & 1.875$\times$1.875$\times$5   & 1.875$\times$1.875$\times$5  & 1.875$\times$1.875$\times$5   \\
\begin{tabular}[c]{@{}l@{}}Recon. Voxel Size\\ - in vivo (mm$^3$)\end{tabular}& 2.00$\times$2.00$\times$5& 2.00$\times$2.00$\times$5& 2.00$\times$2.00$\times$5& 2.00$\times$2.00$\times$5\\
\begin{tabular}[c]{@{}l@{}}FOV - Phantom \\ (mm$^3$)\end{tabular}                                          & 192.5$\times$192.5$\times$100 & 192.5$\times$192.5$\times$100 & 192.5$\times$192.5$\times$90 & 192.5$\times$192.5$\times$100 \\
\begin{tabular}[c]{@{}l@{}}FOV - in vivo \\ (mm$^3$)\end{tabular} & 192.5$\times$160.4$\times$100 & 192.5$\times$160.4$\times$100 & 192.5$\times$160.4$\times$90 & 192.5$\times$160.4$\times$100 \\
\begin{tabular}[c]{@{}l@{}}Scan Time per FA - \\ Phantom (min)\end{tabular}                                       & 5.05                          & 6.87                          & 3.3                          & 5.61                          \\
\begin{tabular}[c]{@{}l@{}}Scan Time per FA - \\ in vivo (min)\end{tabular}             & --& 5.15                          & 2.3                          & 3.88                          \\ \hline
\end{tabular}  
    \label{s-tab:sequence_params}
\end{table}

\begin{table}[H]
\caption{Lower and upper bounds for fitting parameters used in Fat DESPOT$_m$ and Fat DESPOT$_c$.}
\begin{tabular}{lrr}
\hline
Parameter  & Lower Limit   & Upper Limit  \\ \hline
S0         & 0.00001       & $1\times10^{15}$      \\
PDFF (\%)& GC PDFF - 5& GC PDFF +5 \\
$R_2^*$ (s$^{-1}$)& 0             & 1000         \\
$R_{1f}$ (s$^{-1}$)& 0             & 10           \\
$R_{1w}$ (s$^{-1}$)& 0.33           & 10           \\
$\phi_{0f}$ (rad)& 0             & 2$\pi$       \\
$\phi_{0w}$ (rad)& 0             & 2$\pi$       \\ \hline
\end{tabular}
\label{s-tab:bounds}
\end{table}

\begin{figure}[hp]
\begin{center}
 \includegraphics[width=\linewidth]{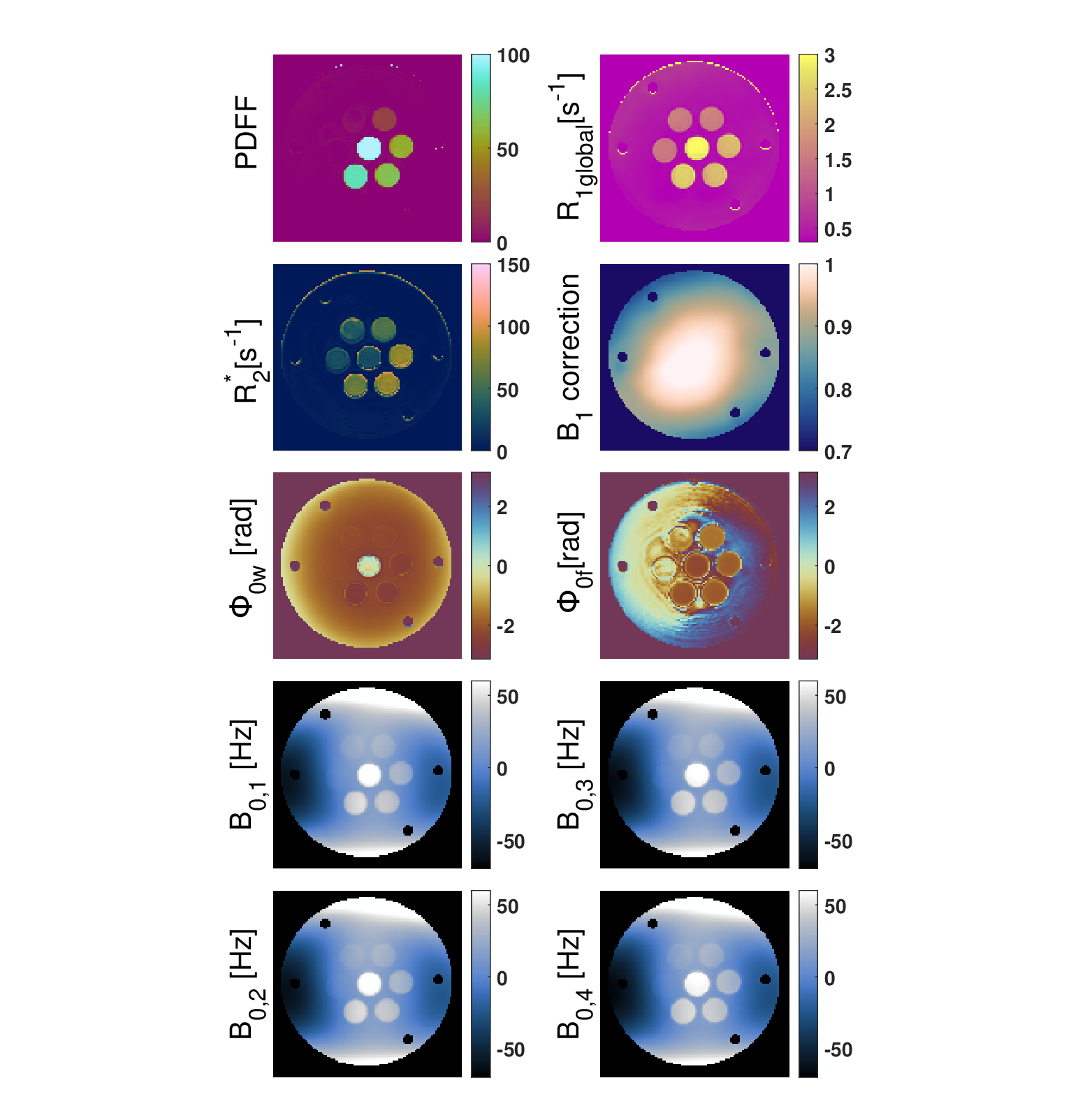}
 \end{center}
  \caption{Examples of initial guess and input parameter maps for the variable fat fraction phantom (8-echo acquisition). The $B_1$ map was obtained from a dual-angle method $B_1$ estimation and $R_{1global}$ map was obtained from a DESPOT$_1$ algorithm on the upper right. All other estimates were obtained using the Graph Cut algorithm.}
  \label{s-fig:initialguess_phantom}
\end{figure}

 \iftoggle{mrm}
 {
 }{
\begin{figure}[hp]
\begin{center}
 \includegraphics[width=\linewidth]{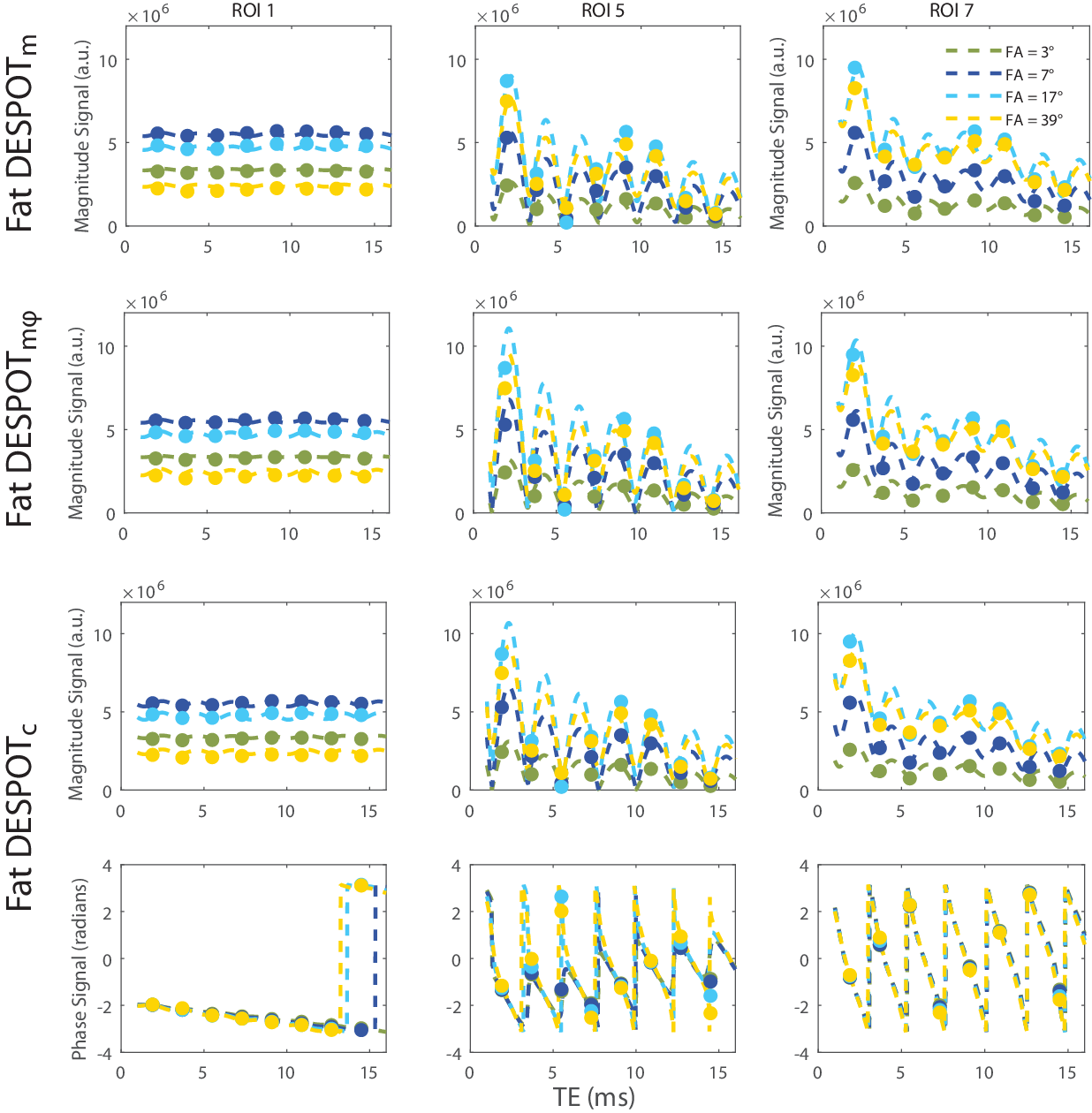}
 \end{center}
  \caption{Examples of voxel-wise fits for the central pixel of 3 ROIs in the variable fat fraction phantom corresponding to nominal fat fractions of 0\% (ROI 1), 50\% (ROI 5), and 100\% (ROI 7). The left column shows the magnitude of the mGRE data (points) and the Fat DESPOT$_m$ fits (dashed line). mGRE data is depicted by points and the Fat DESPOT fits as a dashed lines.}
  \label{s-fig:voxelwise}
\end{figure}
}

\begin{figure}[hp]
\begin{center}
 \includegraphics[width=\linewidth]{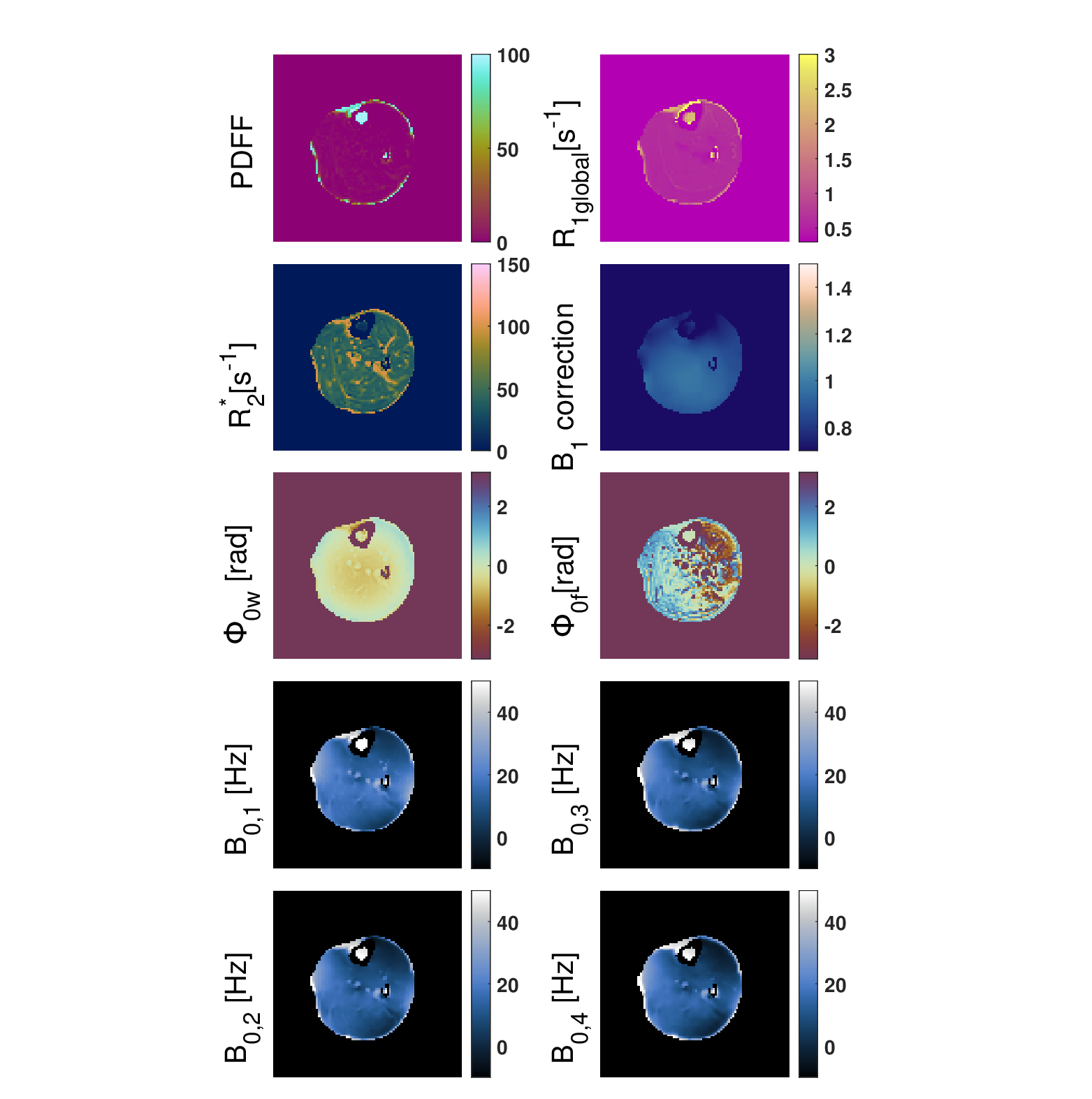}
 \end{center}
  \caption{Examples of initial guess and input parameter maps for the variable fat fraction phantom (8-echo acquisition). The $B_1$ map was obtained from a dual-angle method $B_1$ estimation. The $R_{1global}$ map  obtained from a DESPOT$_1$ algorithm is on the upper right. All other estimates were obtained using the Graph Cut algorithm.}
  \label{s-fig:initialguess_invivo}
\end{figure}
}{
  
}

\end{document}